\documentclass[review]{elsarticle}
\usepackage{geometry}
\usepackage{graphicx}
\usepackage{amssymb}
\usepackage{epstopdf}
\usepackage{hyperref}
\usepackage{color}
\newcommand{\GWt}{GW$_{th}$}

\title{Monitoring Reactor Anti-Neutrinos Using a Plastic Scintillator Detector in a Mobile Laboratory}
\author{J.~Carroll$^1$}
\author{J.~Coleman$^1$}
\author{M.~Lockwood$^1$}
\author{C.~Metelko$^1$}
\author{M.~Murdoch$^1$}
\author{Y.~Schnellbach$^1$}
\author{C.~Touramanis$^1$}
\address{$^1$University of Liverpool, Oliver Lodge Laboratory, Oxford Street, Liverpool, L69 7ZE, United Kingdom}

\author{R.~Mills$^2$}
\address{$^2$National Nuclear Laboratory, Central Laboratory, Sellafield, Seascale, Cumbria, CA20 1PG, United Kingdom}

\author{G.~Davies$^3$}
\author{A.~Roberts$^3$}
\address{$^3$Wylfa Power Station, Cemaes Bay, Gwynedd, LL67 0DH, United Kingdom}

\begin{document}

\begin{abstract}
Technology developed for the T2K electromagnetic calorimeter has been adapted to make a small footprint, reliable, segmented detector to characterise anti-neutrinos emitted by nuclear reactors. The device has been developed and demonstrated by the University of Liverpool and underwent field tests at the Wylfa Magnox Reactor on Anglesey, UK. It was situated in a 20\,ft ISO shipping container, above ground, roughly 60\,m from the 1.5\,\GWt\ reactor core. Based on the design of the T2K Near Detector ECal, the device detects anti-neutrinos through the distinctive delayed coincidence signal of inverse $\beta$-decay interactions using extruded plastic scintillator and Hamamatsu Multi-Pixel Photon Counters.
\end{abstract}

\maketitle

\section{Introduction}
Due to the increasing global energy demand, the number of civilian nuclear power plants is expected to increase in the foreseeable future. Given this development, new technology capable of monitoring the activity of a nuclear reactor becomes increasingly useful. Anti-neutrino detection technology has the potential for non-intrusive monitoring of reactor activity at short to medium distances. In order to allow operation at civilian nuclear power plants, it is crucial for such a device to conform to site safety requirements, be easy to operate and maintain, whilst being compact (a tonne-scale device).

The project presented here aims to provide a reliable, autonomous device for the monitoring of reactor anti-neutrino emissions in safeguards context. A version of such a device was deployed at Wylfa Power Station, Anglesey, UK through the aid of the Department of Energy and Climate Control (DECC) and the Office for Nuclear Regulation (ONR) in the UK. The detector was constructed and commissioned at the University of Liverpool, loaded into a 20\,ft ISO refrigerated shipping container, and shipped to Wylfa via a HIAB truck. The container was situated approximately 60\,m from the centre of the reactor core, outside the inner security barrier of Wylfa reactor building and required only a connection to a power supply for operation. The detector itself is roughly 1\,tonne in mass and utilises technology from the ND280 electromagnetic calorimeter (ECal) of the T2K particle physics experiment~\cite{2011NIMPA.659..106A}. This technology is highly suitable for use on reactor sites thanks to the use of non-toxic materials with a high flash point and an extremely robust mechanical design to ensure that no damage is sustained in transport. The T2K ECal that formed the baseline design for the mechanical structure of this device withstood the magnitude 9 April 2011 earthquake in Japan with minimal damage, demonstrating the robustness of the design.

\section{Reactor Anti-Neutrinos}
Anti-neutrinos are weakly interacting, sub-atomic, particles produced in the $\beta$-decay of nuclei. In operating nuclear reactors with thermal power of order 1\,\GWt, they are produced at a rate of approximately $10^{20}/s$ by the decays of fission daughter nuclei. Both the number of anti-neutrinos produced and their average energy are dependent on the parent fission nuclei. At these anti-neutrino energies the inverse $\beta$-decay (IBD) interaction~\cite{Cowan:1992xc}, $\bar{\nu}_{e} + p \rightarrow e^{+} +n$, is the dominant signal. In the late 1970s, the detection of these reactor anti-neutrinos was first recognised as a means to monitor reactor activity non-intrusively~\cite{institut1978neutrino}. The detection of the positron, followed shortly by detection of a thermal neutron capture gives a distinctive delayed coincidence signal that identifies anti-neutrino interactions. The energy of the detected positron is approximately linearly correlated with the energy of the anti-neutrino, allowing anti-neutrino spectroscopy and core isotope measurements with a detector of sufficient sensitivity.

Anti-neutrino detectors used in particle physics experiments for IBD are usually large (from 10 to multiple 1000\,tonnes) liquid scintillator devices using Photo-Multiplier Tubes (PMT) for readout\cite{Boehm1999166, Apollonio:2002gd, PhysRevLett.90.021802}. Smaller detectors of the order 1\,tonne and constructed of suitably ``safe'' materials, can be placed in close proximity to a reactor whilst being cheap enough for wide scale deployment. Lawrence Livermore National Laboratory developed a prototype, liquid scintillator based, anti-neutrino detector of order 1\,tonne in size, the SONGS1 detector~\cite{2008JPhCS.136b2008B, 2009JoAPsongs}. The detector was deployed at the San Onofre Nuclear Generating Station (SONGS) where reactor anti-neutrinos were used to observe both the operational state of the reactor (on/off) and the fuel burn up during operations. The fuel burn-up manifests as the decrease in anti-neutrino count rate as the fuel cycle progresses and uranium transmutes to plutonium. The observations made by the SONGS1 experiment successfully demonstrated the concept of a small scale reactor anti-neutrino monitoring device to be sound and feasible. This has prompted a number of efforts worldwide~\cite{neos, dayabay, danss, reno, prospect, nucifer}.

These efforts then motivated the IAEA report~\cite{IAEA:neutrinoreport}, noting the ``technical limitations that prevent immediate incorporation''. This report recommended an ideal detector for safeguards should be: of inert construction; non-liquid medium; easy operation; cheap, portable; robust; capable of above-ground operation; ideally with easy deployment in an ISO freight container.

By utilising the isotope dependence of both the anti-neutrino rate and energy~\cite{PhysRevC.83.054615}, it is possible to measure the isotope content of the reactor core as well as its power level over a given time period. As such, anti-neutrino detectors could provide material accounting, non-intrusively, during reactor operation. Due to the weakly interacting nature (i.e. low interaction cross-sections) of anti-neutrino particles, they pass through the containment vessels and other shielding, allowing direct and unshieldable observation of the core. Therefore, a reliable, safe, device could be beneficially deployed as an automated safeguard and accountancy measure, continuously providing detailed information about the reactor core state.

\section{The Liverpool Prototype Detector}
\subsection{Technology}
The device deployed at Wylfa uses neutrino detector technology from the ECal of the T2K experiment designed to precisely measure the energy of electrons and positrons~\cite{2011NIMPA.659..106A, Allan:2013ofa}. In contrast to traditional anti-neutrino detectors the sensitive volume is formed of extruded plastic scintillator bars read out by silicon photo-multipliers, Hamamatsu Multi Pixel Photon Counters (MPPCs). The plastic scintillator is robust, non-toxic, has a high flash point and is cost effective. A wavelength shifting optical fibre is threaded through each bar to guide the light on to the MPPC readout. The MPPCs operate on low voltage, are very compact and incredibly resilient, allowing for a small, transportable detector footprint whilst being easy to operate compared to traditional PMTs. The active detector region includes sheets of gadolinium to allow the detection of thermal neutrons, exploiting its high thermal neutron capture cross-section and its ability to act as a neutron-to-$\gamma$-ray converter~\cite{Kandlakunta01092012}. The electronics readout system is based on the same Field Programmable Gate Array (FPGA) boards used in the T2K experiment with significant firmware adjustments for the detection of inverse $\beta$-decay interactions. An internal veto system is included in the design for the rejection of cosmic ray events and effective operation at ground level.

The anti-neutrino detector contains roughly 1\,tonne of plastic scintillator as the anti-neutrino interaction target and occupies a 1.7\,m $\times$ 1.7\,m footprint instrumented as 1720 individual channels. The plastic scintillator bars measure 40\,mm $\times$ 40\,mm $\times$ 1520\,mm, arranged in a hodoscope layout with alternating scintillator bar direction in the $x$-$y$ plane by 90 degrees each layer. The scintillator planes are interleaved with gadolinium sheets for neutron capture. The 1520\,mm bar length was chosen as it was the shortest used in the T2K ECal. Each scintillator bar uses Kuraray wavelength shifting fibres coupled to the MPPCs to detect the scintillator response. The mechanical support structure ensures correct alignment even during and after transport, allowing assembly off-site and transport of the fully assembled system as a drop-in system, minimising disruption to site operations. The detector is surrounded by 75\,mm of high density polyethylene neutron shielding to reduce background from the fast neutrons produced by the reactor, with 25\,mm of the shielding being boron-loaded to capture thermalised neutrons.

The readout stack is derived from the T2K readout system and consists of 31 Trip-T Front-end Boards (TFB)~\cite{Allan:2013ofa}\ connected to a Readout Merger Module (RMM)~\cite{Allan:2013ofa}\ connected to an IBM-compatible PC running the Maximum Integration Data Acquisition System (MIDAS)~\cite{midasdaq}. Each TFB reads out up to 64 MPPCs and converts the analogue charge signal from the MPPCs using analogue-to-digital converters. In addition, the electronics are adapted to operate on an internal trigger, rather than the external neutrino beam-synchronised pulsed trigger used in T2K. This trigger system fires upon passing a non-localised charge threshold, a signature expected from neutron captures on the gadolinium (see section~\ref{sec:interactions}). Using a FIFO buffer, the detector stores 30\,$\mu$s of data preceding the trigger. The RMM sends the read-out and merged information through a gigabit Ethernet link to the data acquisition PC, recording the data to disk.

\begin{figure}
	\centering
	\includegraphics[width=0.475\textwidth]{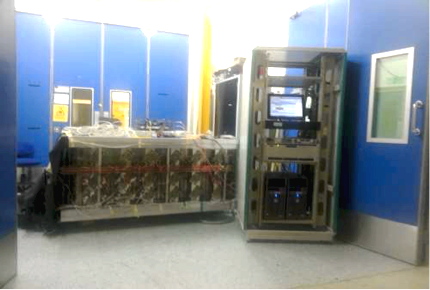}
	\includegraphics[width=0.475\textwidth]{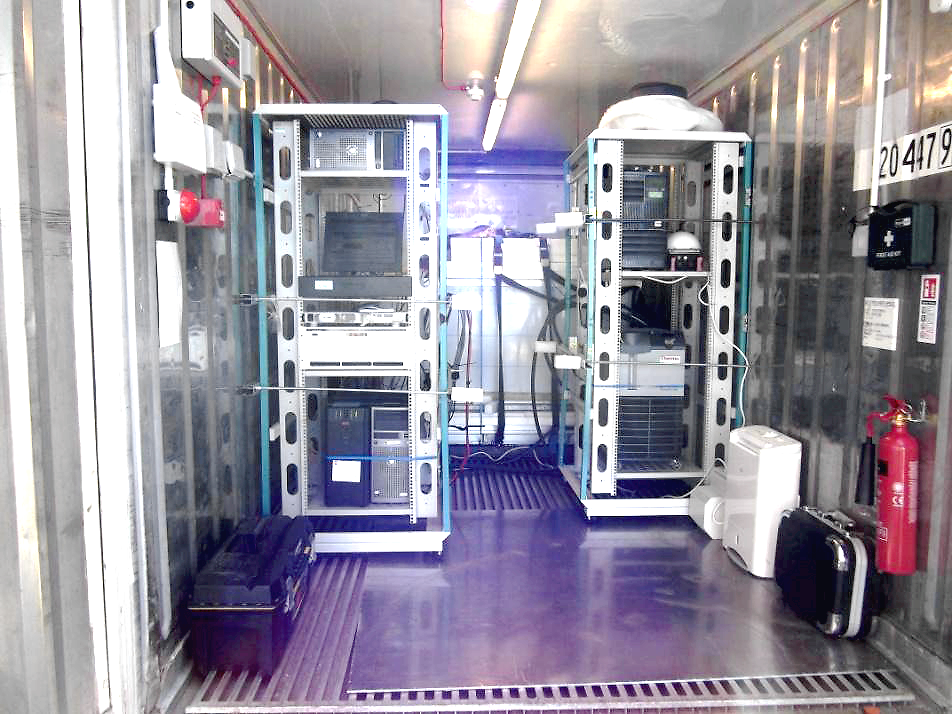}
	\caption{Pictures of the completed anti-neutrino detector. \textit{Left:} at Liverpool. \textit{Right:} Installed into the shipping container.}
	\label{fig:container-inside}
\end{figure}

For ease of transport and deployment, the detector was loaded into a 20\,ft ISO refrigerated shipping container. The shipping container was converted into a mobile laboratory such that all required resources for the operation of the detector were included within, including climate control and power supply. The self-contained unit proved simple to deploy on the reactor site with minimal overheads for the power station staff. The unit requires a 3-phase 420\,V power socket (primarily to power the container's refrigeration unit), and preferably a data connection for remote operations of the device. Pictures of the full detector are shown in figure~\ref{fig:container-inside}.
\subsection{Anti-neutrino interactions in the detector}
\label{sec:interactions}
The plastic scintillator forms the active target region for anti-neutrino inverse $\beta$-decay interactions. These interactions produce a positron and an epi-thermal neutron in the detector. The positron is detected immediately through the light emitted as it passes through the scintillator until it annihilates, resulting in two back-to-back 511\,keV $\gamma$-rays. The majority of the $\gamma$-rays will Compton scatter inside the detector and leave energy deposits close to the current noise threshold and are hence not used in the positron selection. The neutron will be detected later when it captures on a nucleus following a random walk with multiple scatters through the bulk of the detector. Gadolinium is used in the detector as a neutron capture agent due to its high thermal neutron capture cross-section and the release of an 8\,MeV $\gamma$-ray cascade on capture. In the detector the neutron capture is then observed through the Compton Scattering of the emitted $\gamma$-rays, creating a spatially diffuse energy deposit that is strongly correlated in time. The distinct signature of the neutron is used to trigger on events of interest and save them for the data analysis. Anti-neutrino interactions are then selected by looking for the distinctive delayed co-incidence signal of a positron followed by a neutron capture within a given time window. The time elapsed between the positron and the neutron capture event follows an exponential decay curve allowing the subtraction of random coincidence events which display a flat timing distribution. This process is described in detail in section~\ref{sec:bkgsubtraction}.

\subsection{Detector Commissioning and Testing}
\begin{figure}
	\centering
	\includegraphics[width=0.475\textwidth]{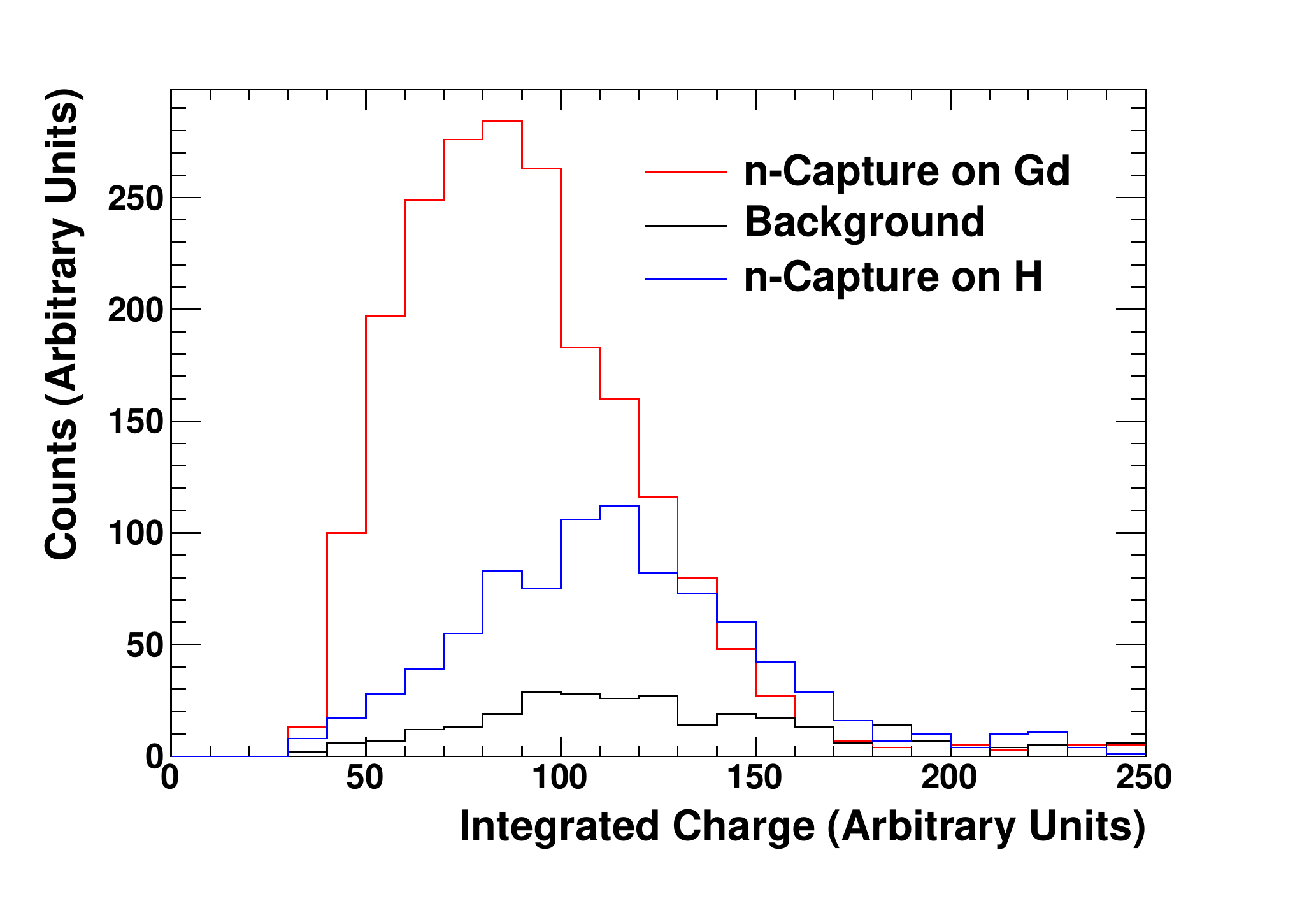}
	\includegraphics[width=0.475\textwidth]{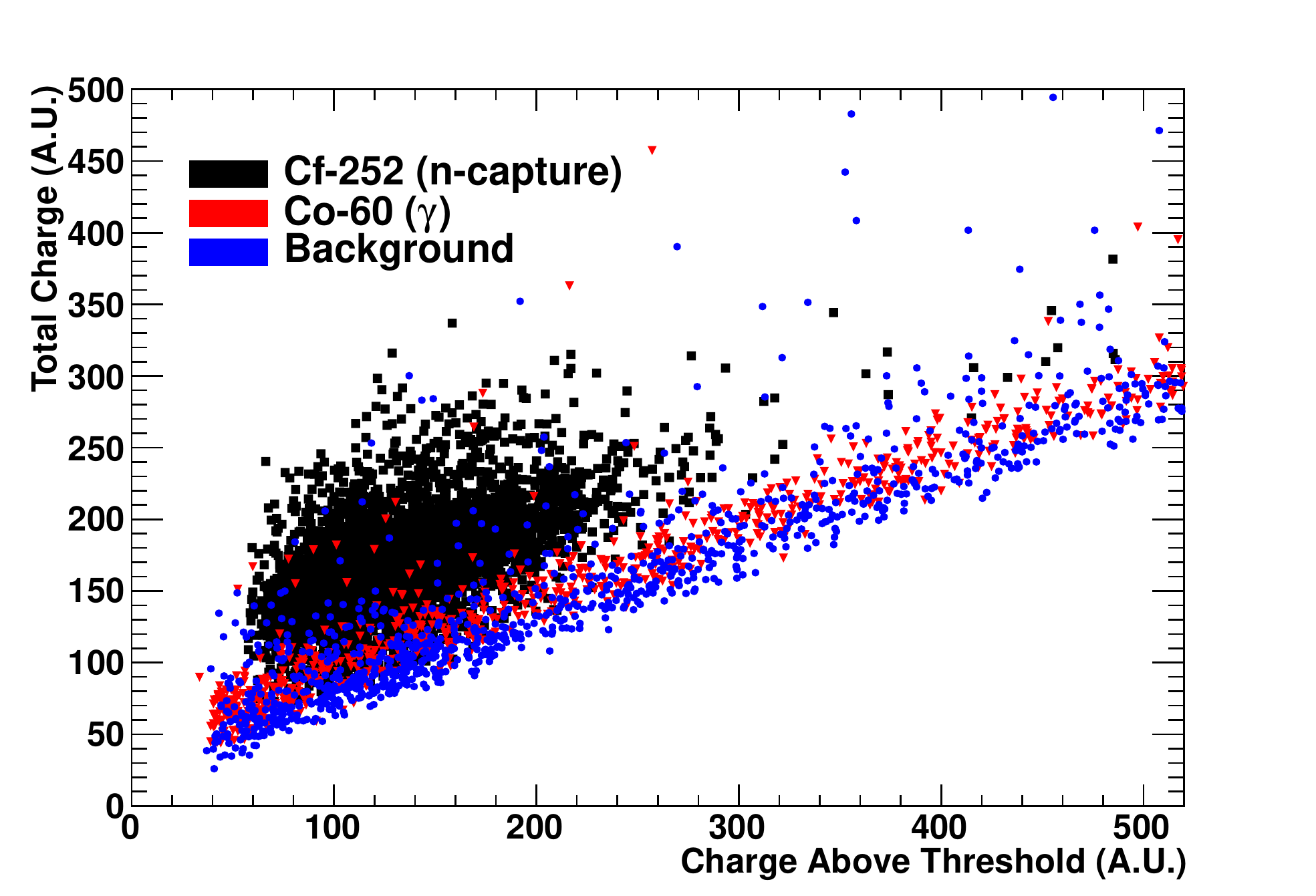}
	\caption{Results of the small scale and full scale source tests. \textit{Left:} A histogram of integrated charge deposit for neutron capture events in a small prototype detector. The integrated area of the curves is proportional to the neutron capture rate. \textit{Right:} The neutron-gamma discrimination abilities for the full scale detector using $^{252}$Cf and $^{60}$Co sources. The x-axis shows integrated charge for individual MPPC hits over a given charge threshold. The y-axis shows total integrated charge.}
	\label{fig:neutrontests}
\end{figure}

A proof-of-concept experiment was built with removable gadolinium inserts to observe the difference between gadolinium and hydrogen neutron captures (hydrogen has a smaller capture cross section and emits a single 2.2\,MeV $\gamma$-ray on neutron capture).  $^{252}$Cf was used as a neutron source. The source was enclosed in lead shielding to attenuate out $>99\%$ of its $\gamma$-ray emissions. Figure~\ref{fig:neutrontests} shows the clear difference in neutron capture rates with and without gadolinium.
A 20-layer prototype was subsequently constructed and tested using the same $^{252}$Cf source with shielding. The 20-layer prototype was also exposed to a $^{60}$Co $\gamma$-ray source for comparison. Figure~\ref{fig:neutrontests} shows the clear neutron-$\gamma$-ray discrimination of the detector.

\begin{figure}
	\centering
	\includegraphics[width=0.9\textwidth]{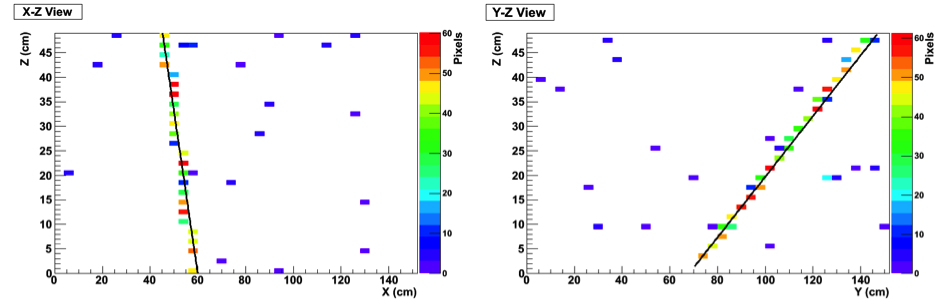}
	\caption{A cosmic ray track passing through the detector. Each plot shows a projection of the track on each face of the detector with each block representing a scintillator bar.}
	\label{fig:cosmic-track}
\end{figure}

For commissioning and testing of the full 49-layer prototype, several data runs were taken selecting cosmic ray muons for data performance checks as well as a cross check for the energy scale calibrations. Muons passing through the detector deposit energy in accordance with the Bethe-Bloch formula. By comparing the observed charge deposit in the detector with the predicted energy deposit, an absolute energy scale calibration can be made. Energy scale calibrations from the radioactive $^{60}$Co source and cosmic ray muons were found to be consistent. Figure~\ref{fig:cosmic-track} shows an event display of a cosmic ray track passing with a fit of its calculated path through the detector. The correct energy response and cosmic ray tracking verify the correct function of the detector. Additionally, this calibration is crucial to future extensions of the analysis to energy spectrum measurements.

\section{Field Tests at Wylfa Power Station}
\subsection{Wylfa Power Station}
The Wylfa site is located on the Isle of Anglesey, North Wales in the U.K. and houses two 1.6\,\GWt\ Magnox reactors which began generating in 1971. Reactor 2 was shut-down permanently in 2012. Reactor 1 continued operation under the novel inter reactor transfer (IRX) programme whereby partially irradiated fuel is transferred from reactor 2. Reactor 1 has been shut-down permanently at the end of 2015.

Magnox reactors are named for the magnesium-oxide cladding on the fuel assemblies. They use natural uranium fuel, carbon dioxide gas cooling and a graphite moderator. Due to the low burn-up of the natural uranium fuel, Magnox reactors were designed to be re-fuelled on load. In terms of the anti-neutrino field tests, this means that the effects of fuel burn-up should be minimal as depleted fuel elements are replaced regularly.

\subsection{Deployment at Wylfa}
\begin{figure}
	\centering
	\includegraphics[width=0.475\textwidth]{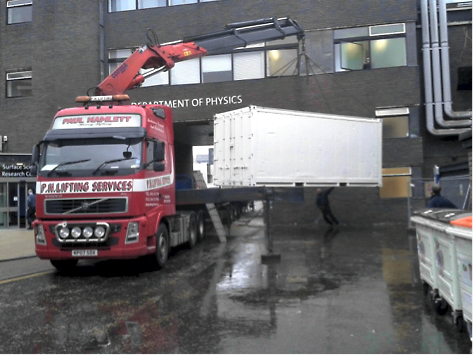}
	\includegraphics[width=0.475\textwidth]{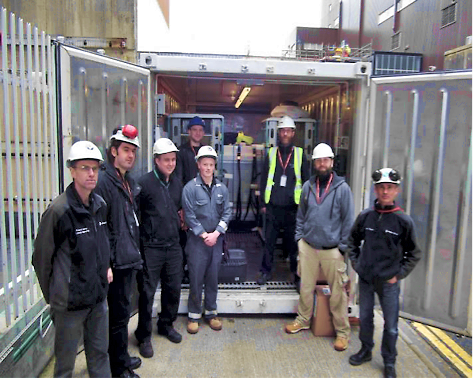}
	\caption{Photographs of the Liverpool detector deployment process. \textit{Left:} Loading of the shipping container onto the flat-bed truck at the University of Liverpool. \textit{Right:} The University of Liverpool team and Wylfa power station staff after the deployment of the container on the reactor site.}
	\label{fig:truckdeploy}
\end{figure}

The University of Liverpool detector was deployed at Wylfa power station for testing starting April 2014, shortly before the start of the final power generating cycle of the Wylfa Magnox powers station. The container was transported by road from University of Liverpool to the Wylfa site. The flat-bed truck used for delivery was an industry standard for transportation of cargo containers. It came with its own HIAB (lifting crane), hence reducing the requirements and burden on the power station for installation. Figure~\ref{fig:truckdeploy} shows the container being loaded on to the transport vehicle at the University of Liverpool. The detector was placed outside the reactor building and next to the turbine halls at a distance of roughly 60\,m from the centre of the reactor core. The detector is outside of internal security barriers surrounding the reactor building, minimising the issues with site access and security clearance at the cost of being further from the reactor core. Figure~\ref{fig:truckdeploy} shows the container and project team on-site after successful deployment.

\subsection{On-site Test Data and Commissioning}
\label{sec:onsite}
On-site checks and commissioning runs were performed after deployment of the detector on-site. Through the analysis of cosmic ray events it was confirmed that no damage had occurred during transit. Minor calibrations were performed to adjust for the change in local environment. The detector then began collecting data before the reactor returned to service. 
\begin{figure}
	\centering
	\includegraphics[width=0.475\textwidth]{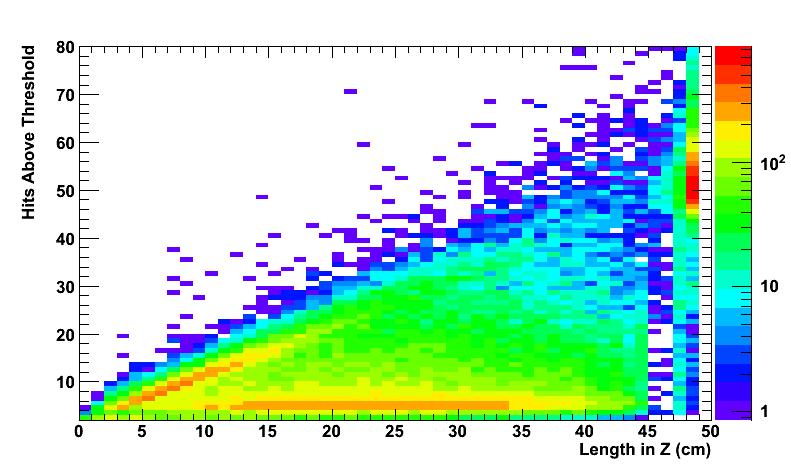}
	\includegraphics[width=0.475\textwidth]{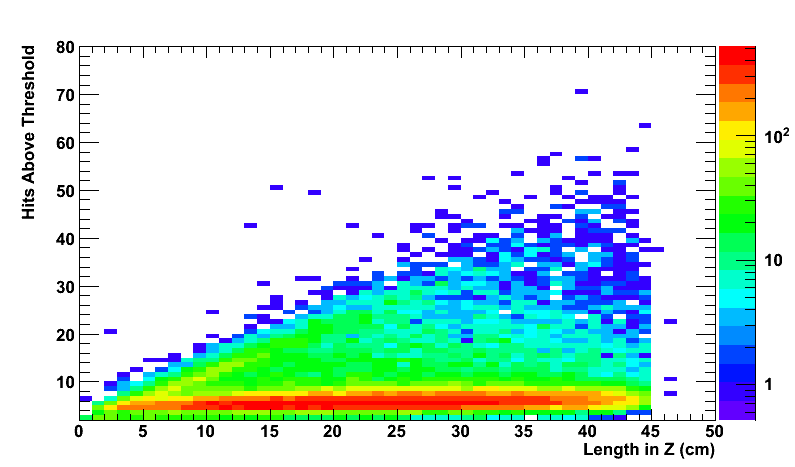}
	\includegraphics[width=0.475\textwidth]{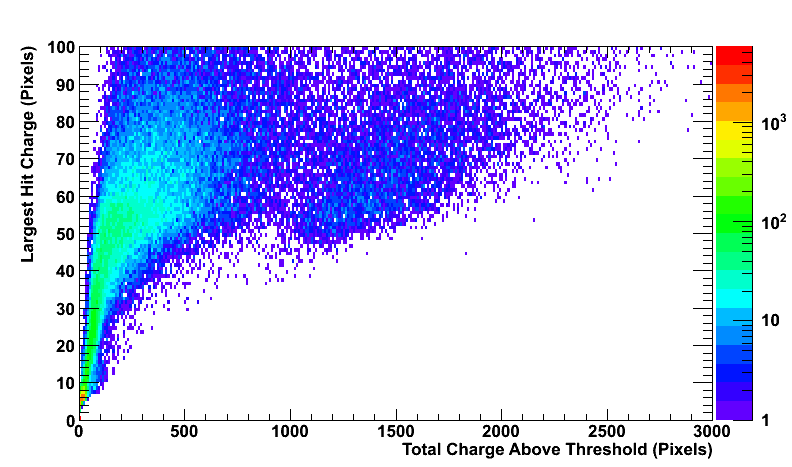}
	\includegraphics[width=0.475\textwidth]{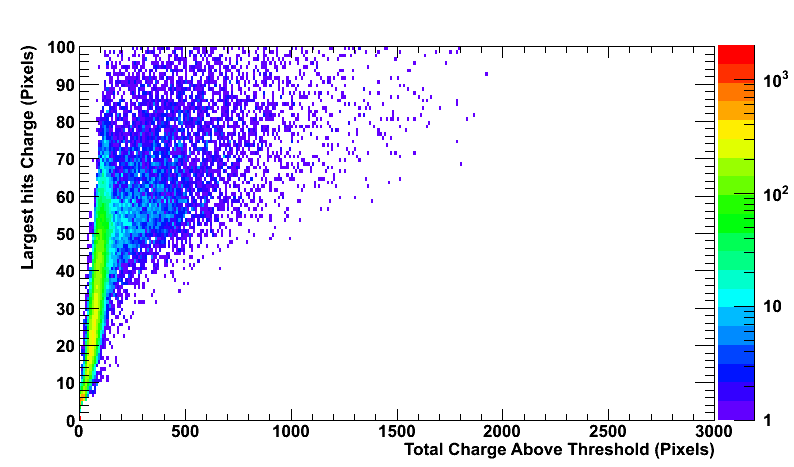}
	\caption{Neutron particle identification plots. \textit{Left:} Data taken at Wylfa before reactor restart combined with cosmic ray data
\textit{Right:} Data taken with a $^{252}$Cf neutron source.
\textit{Top:} Cluster length in \textit{z} plotted against the number of hits above threshold.
\textit{Bottom:} Total charge in detector plotted against the charge of the highest charge hit.}
	\label{fig:neutronPID}
\end{figure}
Further data was taken with a $^{252}$Cf neutron source at the Wylfa site for the development of a neutron identification algorithm in the presence of realistic backgrounds. This data was combined with data from cosmic rays for comparison. Particle identification (PID) comparison plots with the neutron source are shown in Figure~\ref{fig:neutronPID}. The background data contains small clusters of energy deposited by high energy $\gamma$-rays, and cosmic muons that clip the detectors edge, where the spatial size in the detector (x-axis) is roughly equal to the number of scintillator bars hit (y-axis). Cosmic rays that fully penetrate the detector are seen at a cluster length of 49\,cm which is the height of the active region of the detector. From the $^{252}$Cf source data it can be seen that for neutrons there is no correlation between the size of a cluster and the number of hits in the cluster. A series of cuts were developed from these PID plots to select neutron capture candidate events. Candidate neutron capture events are selected at the rate of roughly 1\,Hz in reactor off conditions.

\section{Data Analysis}
\subsection{Event Reconstruction}
The raw data from the detector is processed in the offline data analysis using the ROOT analysis framework~\cite{root}. In order to facilitate high-level candidate event selection (see section~\ref{sec:selection} and~\ref{sec:bkgsubtraction}), the low-level data has to be translated into high-level physics properties. In order to calibrate data for the MPPCs response, dedicated calibration data from a clock-based, non-physics, trigger is used to establish the baseline ``pedestal'' or zero pixel signal shape and rate. Random dark noise within these triggers create a one-pixel peak which is compared to the pedestal to calculate the MPPC gain. These values are then used to convert signals from ADC units to pixels fired equivalent units.

\begin{figure}
	\centering
	\includegraphics[width=0.475\textwidth]{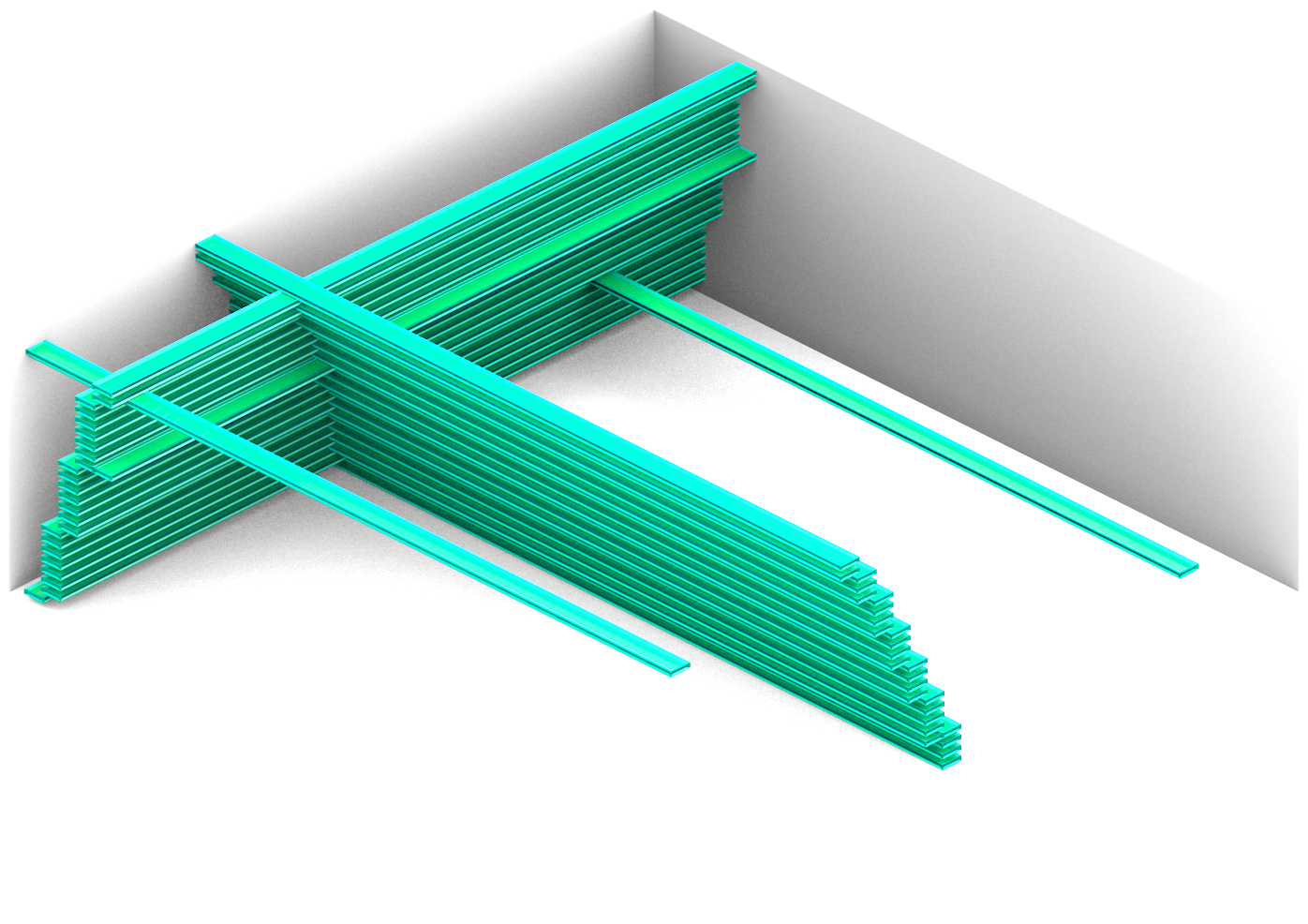}
	\includegraphics[width=0.475\textwidth]{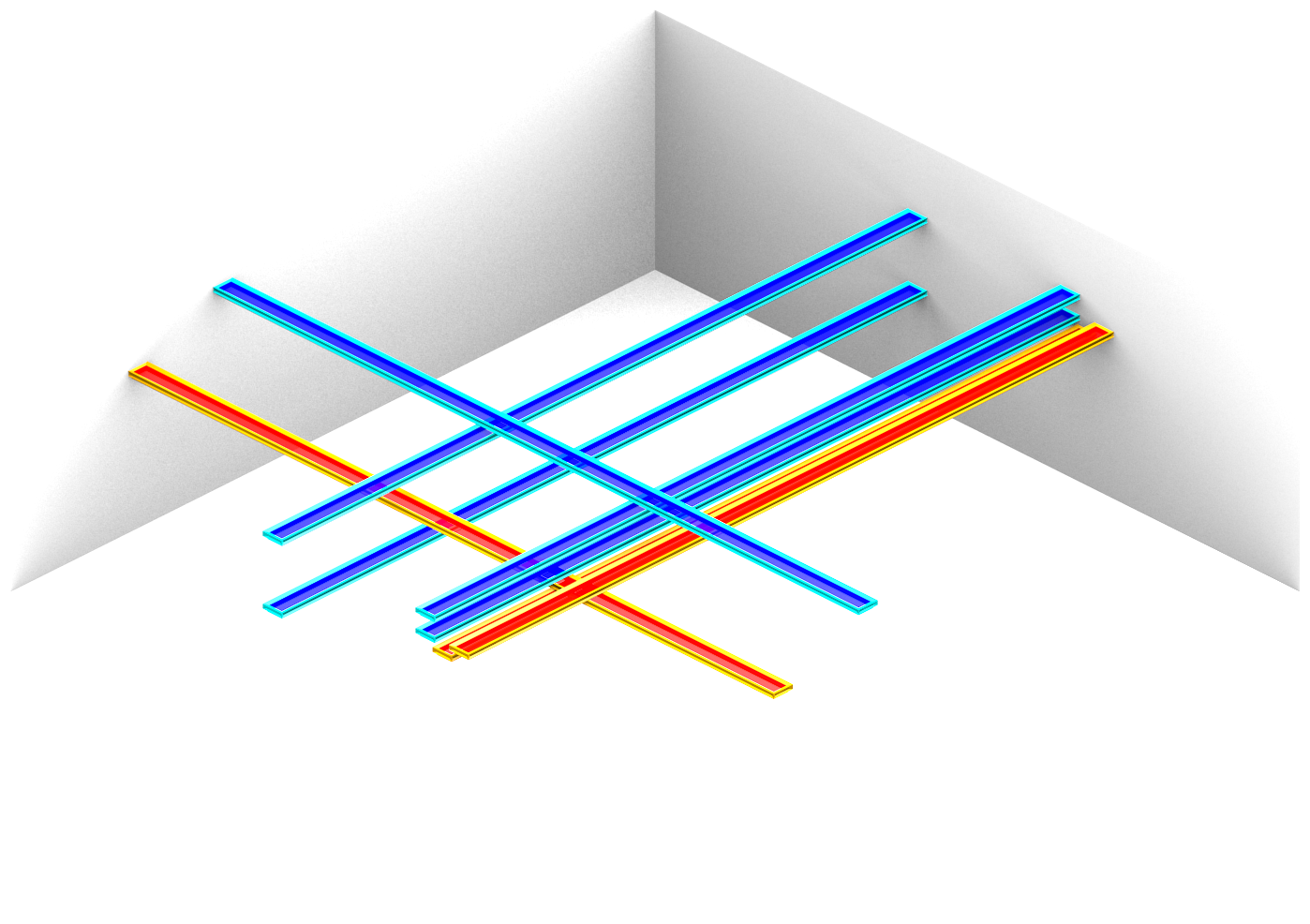}
	\caption{Example event displays of events of interest in the detector. Filled volumes represent hit bars passing the selection criteria for the reconstruction. Green indicates hit bars associated with a cosmic-like track, red indicates a positron-like cluster and blue indicates a neutron-like cluster. \textit{Left:} Cosmic ray event leaving a clear and visible track passing through the entirety of the detector. \textit{Right:} Anti-neutrino candidate event, showing the double coincidence with a short track associated with the positron (red) and the delayed spatially diffuse splash of the neutron (blue).}
	\label{fig:eventdisplay}
\end{figure}

After calibration, nearest neighbour-based reconstruction algorithms are used to identify high-level features of events that can be mapped to physics-related processes. The main features of interest are hit clusters (for positrons and neutrons) and tracks (for cosmic rays). These high-level objects are then used for the following cut-based selection procedure. The topology of events of interest can be seen in figure~\ref{fig:eventdisplay}.

\subsection{Positron Selection and Fast Neutron Rejection}
\label{sec:selection}
The neutron selection is used to select the delayed signal in an inverse $\beta$-decay event as described in section~\ref{sec:interactions}. An equivalent positron selection has been developed to identify the prompt signal component. The positron selection has been partially developed utilising electrons from high energy $\gamma$-ray interactions. Electrons (and positrons) leave a small clustered energy deposit in the detector in the form of a short particle track. The energy range of the positron track is dictated by the energy of the interacting anti-neutrino which informs energy range in which to search for the prompt signal.

Data from the $^{252}$Cf source was also used to develop methods for the rejection of fast neutron backgrounds which produce a correlated time signal similar to that of an anti-neutrino event. The time window for the delayed coincidence search is constrained by the electronics at 30\,$\mu$s in the detector.

Additional position and calibration-derived quantities are used to remove spurious signals. A fiducial volume criterion is implemented to reject events that originated outside of the detector. Furthermore, track fitting algorithms are used to detect events with long tracks associated with cosmic rays which can be used for data quality and calibration measurements.

\subsection{Uncorrelated Background Subtraction}
\label{sec:bkgsubtraction}
\begin{figure}
	\centering
	\includegraphics[width=0.475\textwidth]{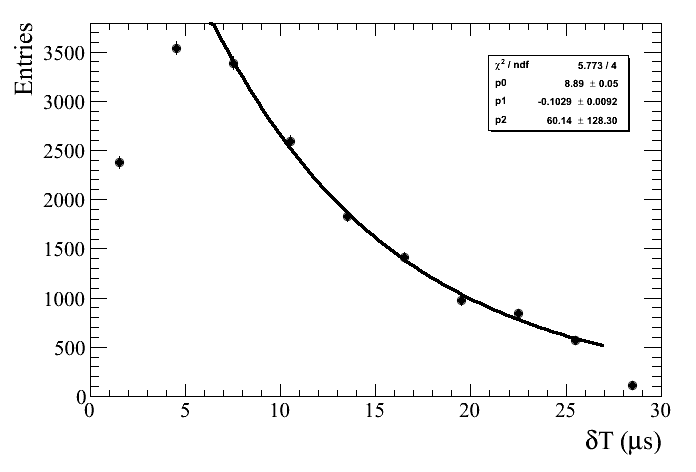}
	\includegraphics[width=0.475\textwidth]{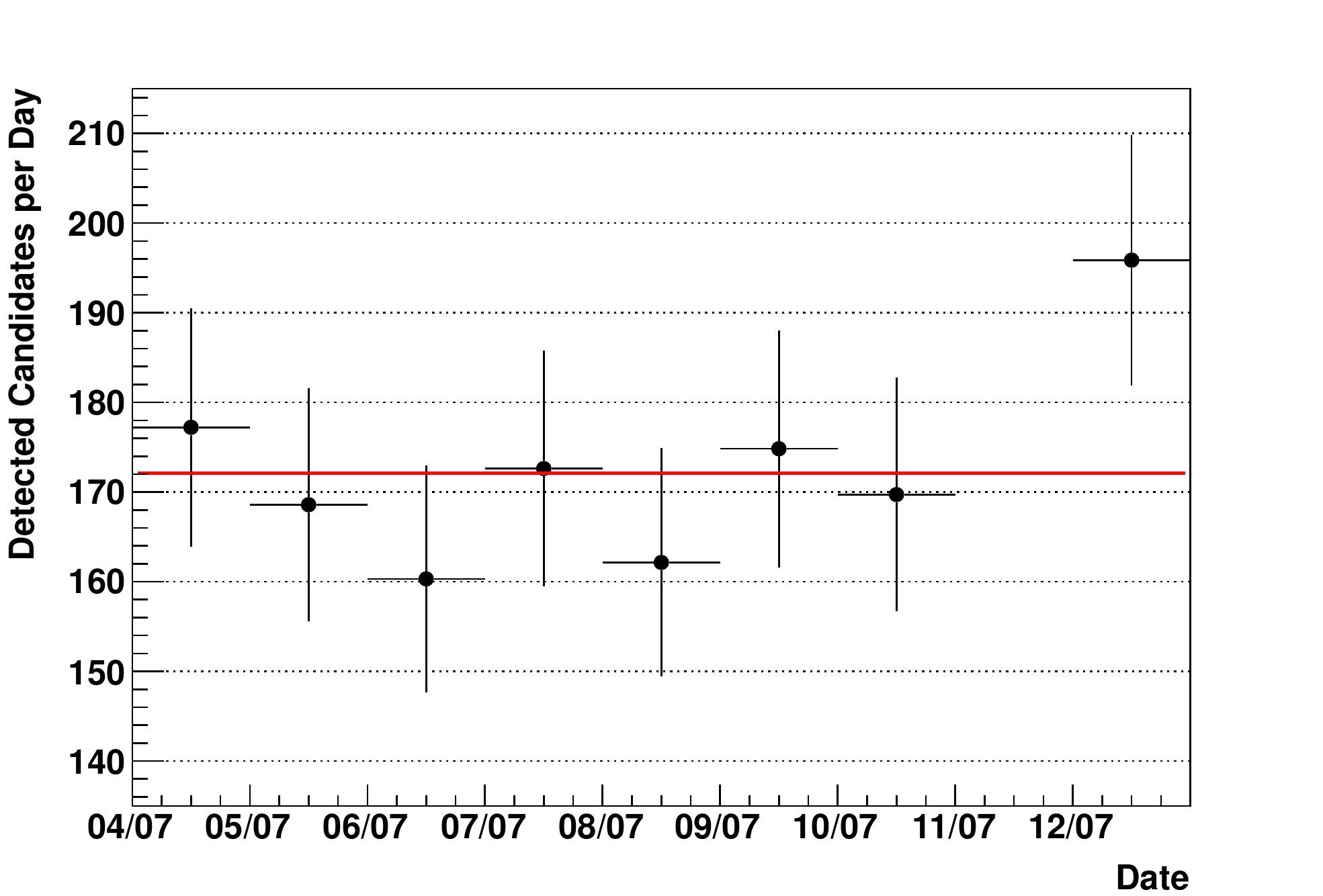}
	\caption{\textit{Left:} The time between delayed and prompt signals for candidate anti-neutrino events during reactor-on data taking after subtraction of uncorrelated backgrounds. Data is fit with an exponential function plus a constant (p2). Detector efficiency effects alter the timing distribution in the first two and the final bins which have been excluded from the fit. \textit{Right:} Daily observed candidate count for the reactor-off period (3 July to 12 July) with a mean of $171.6\pm11.2$ and a corresponding fit of $172.1\pm4.6$ candidate events per day. This rate is used to establish the reactor-off base rate of the detector.}
	\label{fig:bkgsubtime_and_off}
\end{figure}

Uncorrelated background events are when independent positron-like and neutron-like events randomly occur within the 30\,$\mu$s time window of the delayed coincidence search. As the prompt and delayed signals are created independently they show an essentially flat time delay distribution, rather than the exponential decay distribution observed in correlated events. By inverting particle identification event selection criteria, a background sample rich in uncorrelated events is created by rejecting events with particles involved in true coincidence events, leaving a sample of random coincidences. This background-enriched sample is used to establish the rate of random coincidence signals within the search window. By scaling the random coincidence rate to the signal rate a probability that any given event is a random coincidence event is calculated. The random coincidence probability is continuously monitored in the data and the contribution to the anti-neutrino signal candidate events sample is subtracted after the full selection. Figure~\ref{fig:bkgsubtime_and_off} shows the time delay distribution for anti-neutrino candidate events after subtraction of uncorrelated backgrounds. The data is fit with an exponential decay plus a constant. The constant is consistent with zero, indicating that the majority of the uncorrelated background has been subtracted.

\section{Results}
\subsection{Reactor Turn-On Observation}
Reactor 1 at the Wylfa Magnox plant was restarted mid-month In July 2014. This gradual ramp-up began with operation at ca. 0.9\,\GWt, followed by an increase to ca. 1.6\,\GWt\ towards the end of the month. The reactor-off period was used to establish a baseline background rate, resulting in a mean of $171.6\pm11.2$. The reactor-off data were fitted with a constant, resulting in a fit value of $172.1\pm4.6$, showing good agreement with both the mean, and a constant stable background rate. The reactor-off data and the corresponding fit are shown in figure~\ref{fig:bkgsubtime_and_off}.

\begin{figure}
	\centering
	\includegraphics[width=0.9\textwidth]{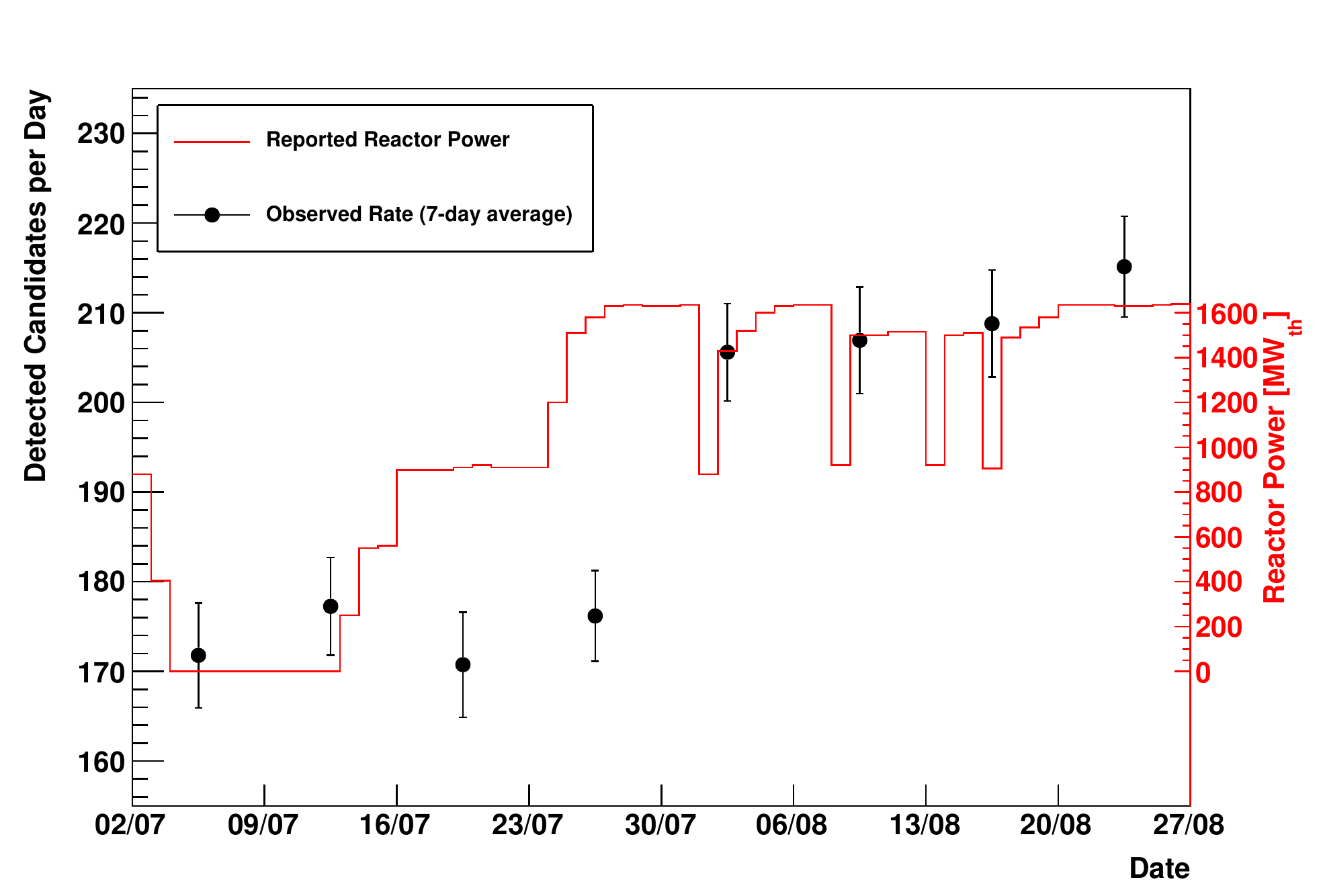}
	\caption{Daily count rates of detected anti-neutrino candidates averaged over 7-day intervals for the Wylfa power plant switch-on and the reactor thermal power reported by the plant operator. Error bars indicate the statistical uncertainty only.}
	\label{fig:onoff}
\end{figure}

Shortly after the off-period, a neutron source was brought for tests as described in section~\ref{sec:onsite} on 16 July resulting in a thermal shift and miscalibration for regular operation causing a loss in detector efficiency until 28 July. After recalibration and recovery, the detector continued to take data with the reactor at full power. During the nominally calibrated full power period (29 July to 21 August), an increased candidate event count of $203.7\pm19.6$\ candidates per day was observed. Shown in figure~\ref{fig:onoff} are the daily candidate counts averaged in 7-day intervals versus the reactor thermal power reported by the site operators, showing the clear increase in candidates with the operating reactor, even during the half-power period and the period of lowered detector efficiency. Subsequent to the observation of the reactor turn on, the detector was reconfigured to investigate alternative running conditions.

\subsection{Statistical Significance}
\begin{figure}
	\centering
	\includegraphics[width=0.475\textwidth]{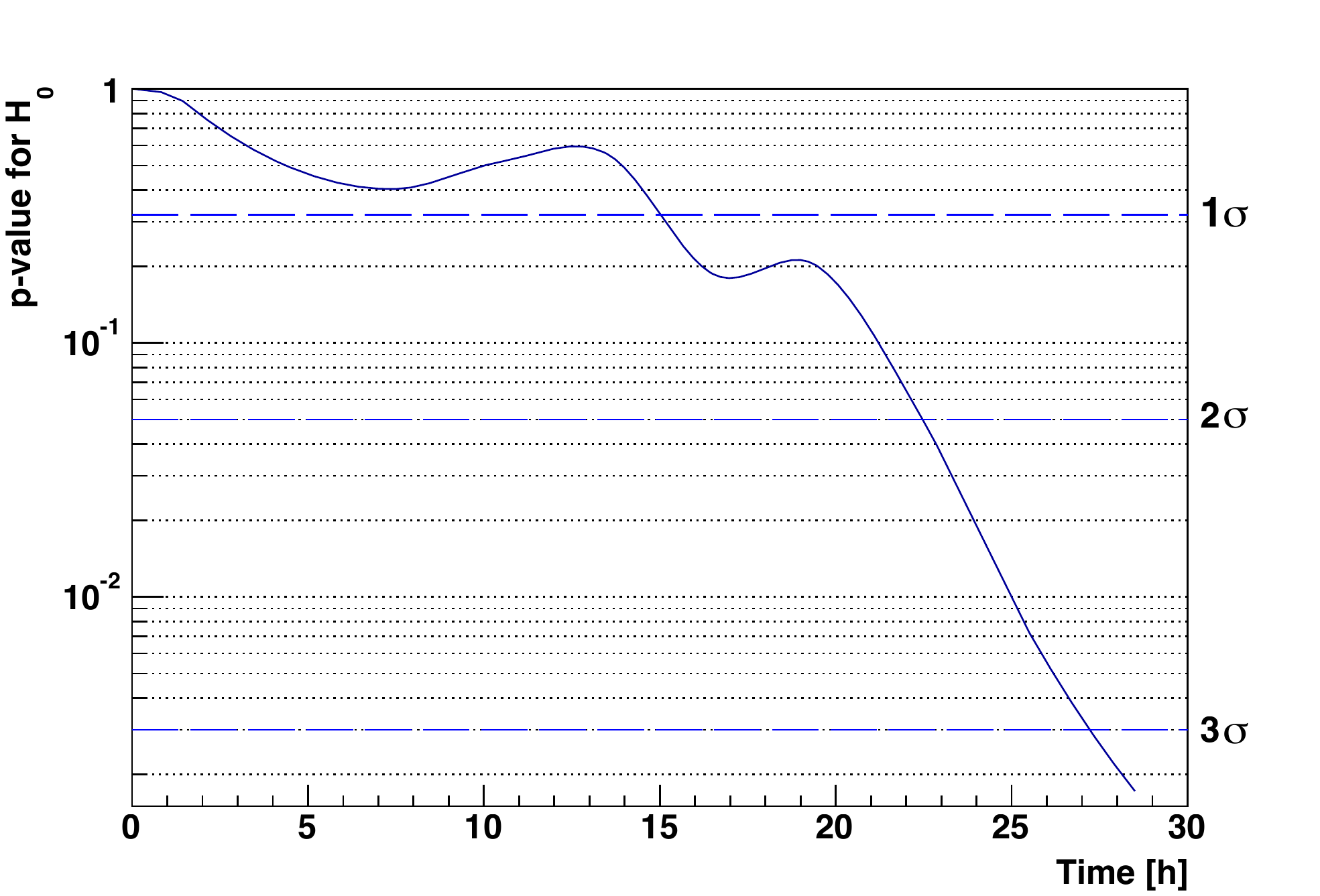}
	\includegraphics[width=0.475\textwidth]{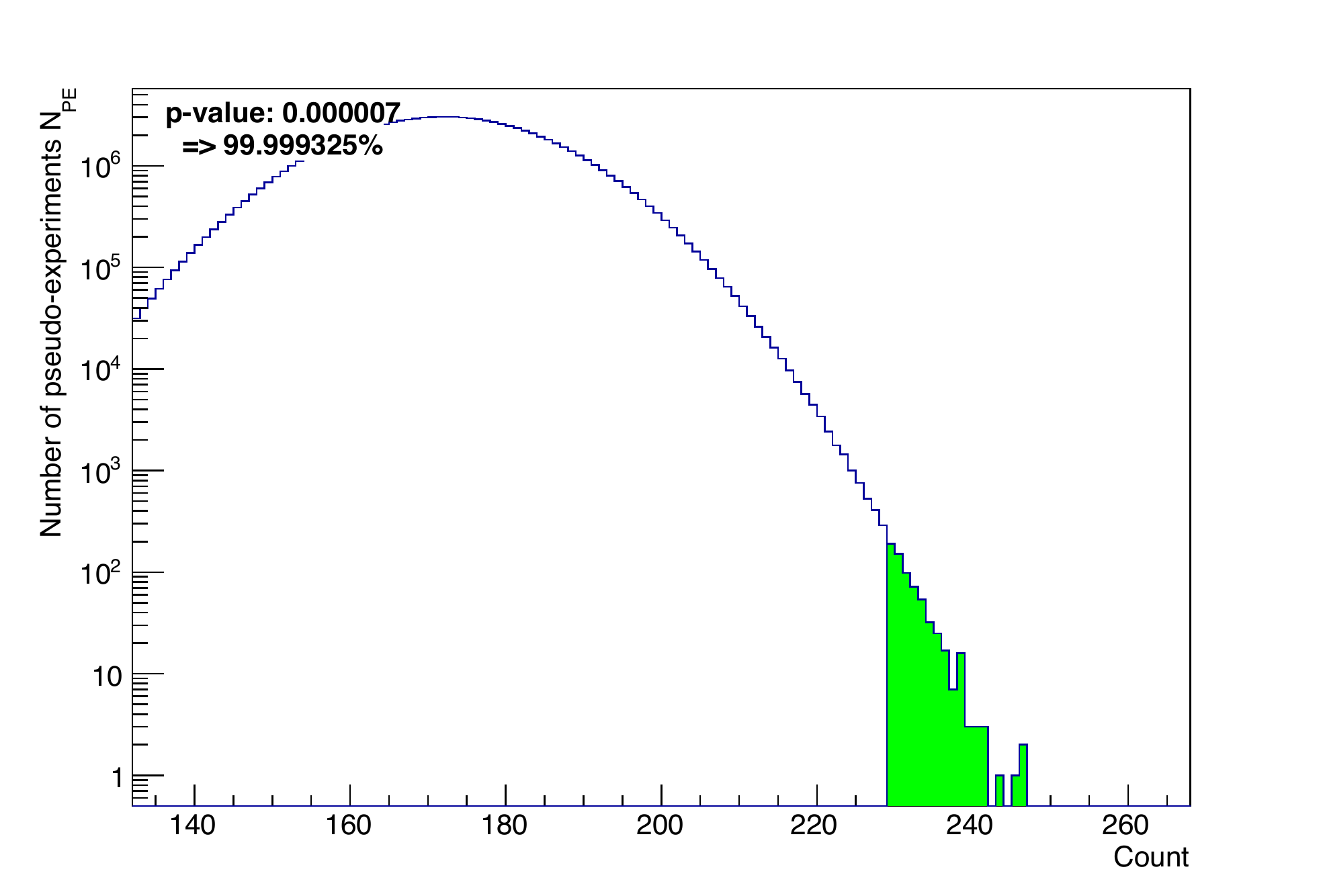}
	\caption{\textit{Left:} Null hypothesis $p$-values for the cumulative 3-hourly candidate count during the reactor full power period, starting at 29 July. The dashed lines indicate 1, 2 and 3 $\sigma$ significance thresholds for candidates in excess of the null hypothesis. \textit{Right:} Pseudo-experiment distribution for the expected average observation of 228.7 events at 45\,m distance with 100,000,000 pseudo-experiments, with an expected excess of over $4\sigma$ significance after a day of operation.}
	\label{fig:pval}
\end{figure}

In order to qualify the statistical significance of the observed excess in candidates, the reactor-off observation is used to form the null hypothesis. Since the Poisson error ($\sqrt{171.6}=13.1$) for the observed reactor off rate shows good agreement with the observed standard deviation ($\pm10.3$), the expected candidate count uncertainty is assumed to be purely statistical. The cumulative expected candidate count is then calculated in 3\,hour intervals and compared to the observed cumulative candidate count to calculate the corresponding $p$-values in favour of the null hypothesis (reactor-off).
The clear downward trend in $p$-values shown in figure~\ref{fig:pval} indicates an accumulating excess of event candidates. The excess reaches $1 \sigma$\ significance after 15\,h, $2 \sigma$\ significance after 24\,h and $3 \sigma$\ significance after 27\,h. As the anti-neutrino flux falls with the inverse square of the distance from the core, moving the detector to 45\,m distance is expected to raise the average candidate rate to ca. 228.7 per day. In a toy experiment, the expected reactor-on value tested against the null hypothesis, as shown in figure~\ref{fig:pval} using pseudo-experiments. This test results in a $p$-value of $6.75\times10^{-6}$ within 24\,h ($4.4\sigma$), not accounting for the statistical spread of the observed candidate count. Hence, at 45\,m distance, a $3\sigma$ excess is expected to be seen within a single day. 

\section{Summary and Outlook}
An anti-neutrino detector for safeguards purposes has been constructed and commissioned at the University of Liverpool based on the technology used for the ECal of the T2K experiment. The detector design uses only safe and robust materials in order to be as safeguards- and operations-compliant as possible. Commissioning data shows the detector to be highly effective at neutron-$\gamma$-ray discrimination due to the use of a gadolinium neutron absorber. Combined with the ability to detect positrons, the anti-neutrino detector is well suited to the observation of inverse $\beta$-decay interactions. The detector has been deployed in an ISO container at the Wylfa power station reactor site in the UK for field testing. Initial results from analysis of the data and selection of candidate anti-neutrino events have shown the ability to observe statistically significant excess (3$\sigma$) of anti-neutrino-like candidate events during active reactor periods after less than 2 days of operation at 60\,m distance. An series of upgrades to the detector based on the experience at Wylfa nuclear power station are currently being implemented in conjunction with John Caunt Scientific Ltd, supported by a grant from Innovate-UK.

\section*{Acknowledgements}
This work was supported of the STFC Innovations Partnership Scheme (IPS), the Royal Society of Edinburgh, the Royal Society.  We are grateful for the support and contributions of the Particle Physics group and the mechanical workshop from the Physics Department at the University of Liverpool. We would like to thank the members T2K-UK for their advice and the loan of relevant components.  The authors are also grateful for the support and effort from the staff at Wylfa Power Station. The necessary introductions for site access were facilitated through the Department of Energy and Climate Control (DECC) and the Office for Nuclear Regulation in response to approach from the UK Support Programme to the IAEA (UKSP). The UKSP is funded by DECC to provide technical support to the IAEA Department of Safeguards.

\end{document}